\documentclass[aps,floatfix,twocolumn,final,prl,showpacs]{revtex4}

\usepackage{epsfig}
\usepackage{graphicx}
\usepackage[ansinew]{inputenc}
\usepackage{array}
\usepackage{hyperref}
\usepackage{tabularx}
\newcolumntype{C}[1]{>{\centering\let\newline\\\arraybackslash\hspace{0pt}}m{#1}}

\begin{document}

\preprint{APS/123-QED}
\title{Metallic islands in the Kondo insulator SmB$_{6}$}

\author{J. C. Souza$^{1,2}$, P. F. S. Rosa$^{3}$, J. Sichelschmidt$^{2}$, M. Carlone$^{4}$, P. A. Venegas$^{5}$, M. O. Malcolms$^{1}$, P. M. Menegasso$^{1}$, R. R. Urbano$^{1}$, Z. Fisk$^{6}$, and P. G. Pagliuso$^{1}$}

\affiliation{$^{1}$Instituto de F\'isica \lq\lq Gleb Wataghin\rq\rq,
UNICAMP, 13083-859, Campinas, SP, Brazil\\
$^{2}$Max Planck Institute for Chemical Physics of Solids, D-01187 Dresden, Germany\\
$^{3}$Los Alamos National Laboratory, Los Alamos, New Mexico 87545, USA\\
$^{4}$Departamento de F\'isica, Faculdade de Ci\^encias, UNESP, C.P. 473, 17033-360, Bauru, SP, Brazil\\
$^{5}$POSMAT, Faculdade de Ci\^encias, UNESP, C.P. 473, 17033-360, Bauru, SP, Brazil\\
$^{6}$Department of Physics and Astronomy, University of California, Irvine, California 92697,USA}


\date{\today}

\begin{abstract}
The predicted interplay between Kondo physics and non-trivial topology in SmB$_{6}$ has stimulated many experimental reports, some of which are in apparent contradiction. The origin of the dispute may lie on the fragility of the Kondo insulating phase in the presence of Sm vacancies (Kondo holes) and/or natural impurities, such as Gd$^{3+}$. In this work, we locally investigated this fragility for Al-flux grown Sm$_{1-x}$Gd$_{x}$B$_{6}$ single crystals (0 $\leq$ $x$ $\leq$ 0.02) by combining electron spin resonance (ESR) and complementary bulk measurements. The Gd$^{3+}$ ESR spectra in a highly dilute regime ($x$ $\sim 0.0004$) display the features of an insulating cubic environment. Remarkably, a metallic ESR lineshape is observed for more concentrated samples ($x$ $\geq$ 0.004), even though these systems are still in a reasonably dilute regime and show insulating $dc$ electrical resistivity. Our data indicate that the Kondo insulating state is destroyed locally around impurities before a global percolation occurs. This result not only explains the discrepancy between $dc$ and $ac$ conductivity, but also provides a scenario to explain the presence of quantum oscillations in magnetization in the absence of quantum oscillations in electrical resistivity.
\end{abstract}

\maketitle

\section{\label{sec:intro}I. Introduction}

The Kondo insulator SmB$_{6}$ has attracted a lot of interest during almost half a century due to numerous puzzling properties such as the physics of the hybridization gap, the mixed-valence ground state, nonzero specific heat at low temperatures, the crystalline electrical field (CEF) ground state and the saturation in the resistivity under $T$ $\approx$ 4 K \cite{dzero2016topological}. The interest was renewed since the prediction that SmB$_{6}$ is a topological Kondo insulator (TKI) \cite{dzero2012theory,alexandrov2013cubic}. Many experimental results support a TKI phase \cite{sundermann2018fourf,kim2013surface,kim2014topological,neupane2013surface,jiang2013observation,xu2014direct,wolgast2013low}, but this classification remains a matter of contention \cite{hlawenka2018samarium}. Conflicting experimental results of SmB$_{6}$ further complicate classification schemes.

For instance, quantum oscillations in magnetization are observed in floating-zone grown samples, whereas they are absent in aluminum flux grown ones \cite{li2014two,tan2015unconventional,thomas2018quantum}. Electrical conductivity measurements also appear conflicting. A bulk activated insulating behavior is obtained in $dc$ electrical resistance measurements \cite{eo2018robustness}, while $ac$ conductivity measurements show localized states with conductivities orders of magnitude higher than the $dc$ measurements \cite{gorshunov1999low,laurita2016anomalous,laurita2019impurities}. 

These discrepancies naturally invite a discussion regarding the role of impurities and defects in SmB$_{6}$. Raman spectroscopy measurements explored the effect of Sm vacancies and argued for a breakdown of the Kondo insulating phase for small amounts of Sm vacancies \cite{valentine2016breakdown,valentine2018effect}. However, Corbino disk $dc$ resistance measurements have shown that Sm vacancies do not affect the thermally activated bulk behavior of flux-grown SmB$_{6}$, indicating that the bulk may be immune to disorder in the $dc$ limit \cite{eo2018robustness}. The effect of Sm vacancies acting as \lq\lq Kondo holes\rq\rq \cite{hamidian2011how}, i.e., an isolated nonmagnetic impurity in a Kondo lattice \cite{sollie1991simple,schlottmann1996metal,figgins2011defects,lu2013quantum}, also has consequences for the formation of a possible TKI phase \cite{valentine2016breakdown}. On one hand, low-energy spin excitons could destroy the protection of the gapless surface states \cite{arab2016effects}. On the other hand, Kondo holes are argued to enable quasiparticle interference patterns that reveal the heavy surface states in recent scanning tunneling spectroscopy (STS) measurements \cite{pirie2018imaging}. Furthermore, STS results showed that Gd$^{3+}$ impurities in SmB$_{6}$ destroy the surface states locally and the effect percolates for $x$ $\geq$ 0.03 \cite{jiao2018magnetic}.

An outstanding question is whether magnetic impurities such as Gd$^{3+}$, which is a natural impurity in Sm, could also display similar effects as the Sm vacancies on the hybridization gap of SmB$_{6}$. In order to properly address this question, the use of a microscopic technique that locally probes the effects of Gd$^{3+}$ in SmB$_{6}$ is highly desirable. Recent experimental results by Fuhrman $et$ $al.$ were explained assuming the possibility of a dynamic screening of localized Gd$^{3+}$ moments, which is unexpected due to the Gd$^{3+}$ 4$f$$^{7}$ electronic configuration, which carries no orbital moment and is particularly stable \cite{fuhrman2018screened,fuhrman2018magnetic}. 

The Gd$^{3+}$ spin is a standard probe in electron spin resonance (ESR) experiments; however, to the best of our knowledge, there is no report of ESR in highly dilute samples ($x$ $\leq$ 0.0004), except the observation by G. Wiese $et$ $al.$ \cite{wiese1990possible} of an anomalous spectrum of Gd$^{2+}$ for concentrations down to 200 parts per million. This spectrum most likely originates from the resonance of a Gd$^{2+}$ ion within an oxide impurity phase on the crystal surface \cite{souza2019anomalous}. Other reports are not reproducible \cite{kojima1978electron,kunii1985electron}, which begs for a revisited experimental investigation. 

Here we present ESR and complementary macroscopic measurements of high-quality flux-grown Sm$_{1-x}$Gd$_{x}$B$_{6}$ single crystals with nominal concentrations $x$ = 0, 0.0004, 0.004 and 0.02. From magnetic susceptibility data and using the ESR-determined actual Gd$^{3+}$ concentrations, we extracted a Gd$^{3+}$ magnetic moment that is close to the theoretically expected value. In the highly dilute regime ($x$ = 0.0004) at $T$ = 4 K, the Gd$^{3+}$ ESR shows seven symmetrical Lorentzian lines typical of a cubic insulating environment. Increasing the Gd$^{3+}$ concentration to $x$ = 0.004, a single asymmetrical line shape appears, which is characteristic for metallic samples where the microwave skin depth is smaller than the sample size \cite{feher1955electron,dyson1955electron,barnes1981theory}. These results suggest that Gd$^{3+}$ ions could close the hybridization gap locally while the resultant metallic islands do not affect the global hybridization gap in this concentration range.

\section{\label{sec:experiment}II. Methods}

Single crystalline samples of Sm$_{1-x}$Gd$_{x}$B$_{6}$ were synthesized by Al-flux grown technique as described elsewhere \cite{eo2018robustness}. The samples had a typical size of $\sim$ 700 $\mu$m width, 300 $\mu$m height and 120 $\mu$m thickness. Magnetic susceptibility measurements were carried out in SQUID and SQUID-VSM magnetometers. Specific heat measurements were performed in a commercial small-mass calorimeter system. Electrical resistivity was measured using a standard four-point technique with a $dc$ bridge. ESR measurements were performed on single crystals in a X-band ($\nu$ $\cong$ 9.4 GHz) spectrometer equipped with a goniometer and a He-flow cryostat in the temperature range of 2.6 K $\leq$ $T$ $\leq$ 40 K at very low power of $P$ = 0.21 mW. In order to calibrate the ESR intensity in our spectrometer to estimate the actual Gd$^{3+}$ concentrations, we used a standard weak pitch sample with 0.79$\cdot$10$^{14}$ spins/cm$^{3}$ and a $g$-value $g_{ref}$ = 2.002(1). The samples were etched before the ESR measurements in a mixture of hydrochloric and nitric acids in a proportion of 3:1 to remove any possible impurities on the surface of the crystals due to Al flux. We did not polish any crystals in this study. The mass of the studied samples ranged from 0.3 mg to 4 mg. We performed our experiments in 20 different crystals from five different batches and found no sample dependence. The concentration $x$ used below refers to the nominal concentration value.

\section{\label{sec:resultsanddiscussion}III. Results and Discussion}

\begin{figure}[!ht]
\includegraphics[width=0.8\columnwidth]{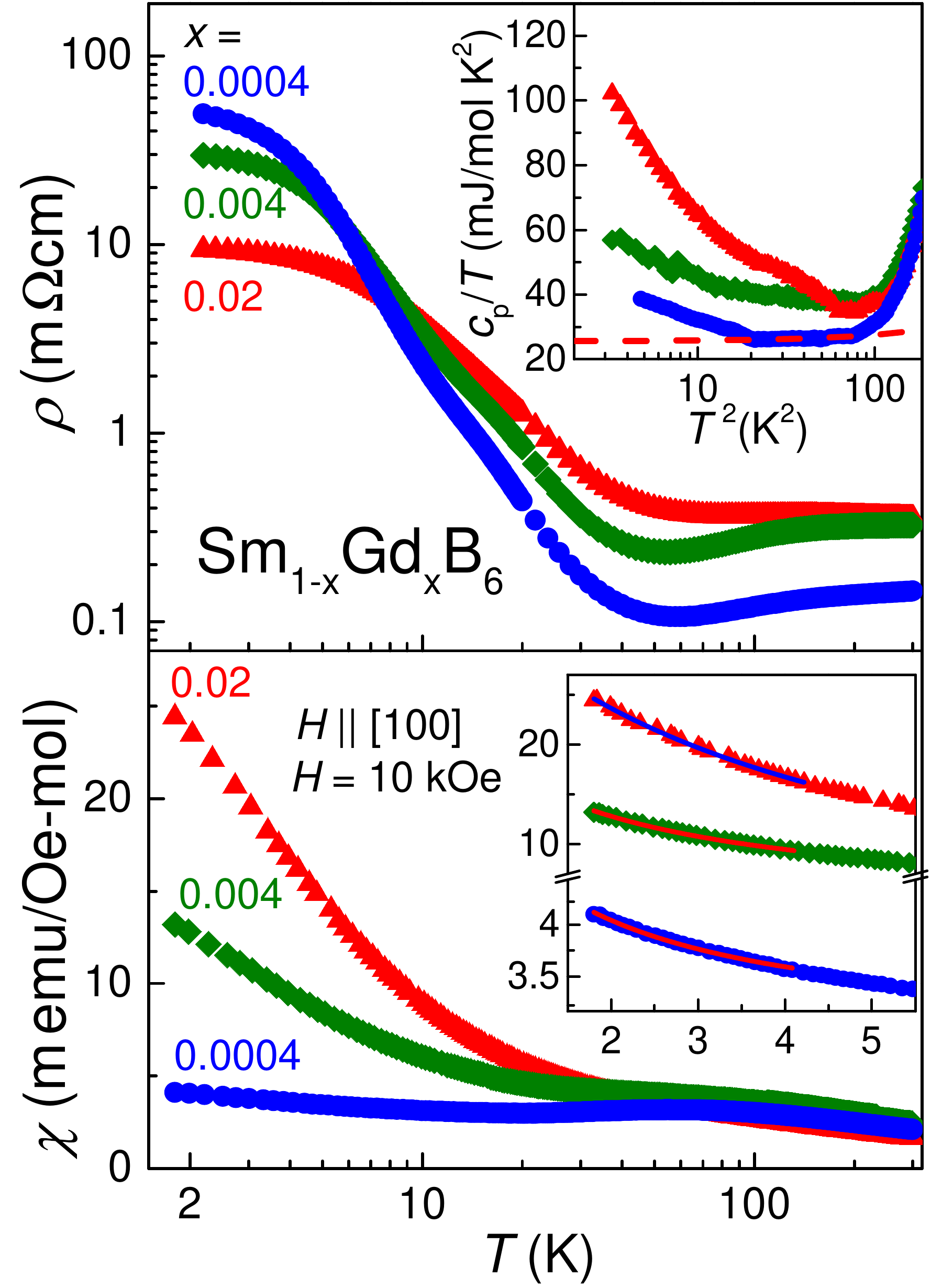}
\caption{Temperature dependencies of $dc$ electrical resistivity $\rho$, specific heat $c_{p}$ and $dc$ magnetic susceptibility $\chi$ for Sm$_{1-x}$Gd$_{x}$B$_{6}$ with $x$ as indicated. For $\chi$, the magnetic field $H$ = 10 kOe is applied parallel to the [100] direction. Lower inset shows $\chi$ at low temperatures. For $\chi$, the solid lines are the best fits obtained considering a Curie-Weiss plus a Pauli $T$-independent term. For $c_{p}$ the red dashed line shows, for $x$ = 0.0004, how we estimated the $\gamma$ value.}
\label{Fig1}
\end{figure}

Figure \ref{Fig1} summarizes the bulk macroscopic properties of Sm$_{1-x}$Gd$_{x}$B$_{6}$ with $x$ = 0.0004, 0.004 and 0.02. The top panel of Fig. \ref{Fig1} shows the resistivity as a function of temperature. At low temperatures, the three concentrations show a saturation in the resistivity, consistent with the undoped compound \cite{kim2014topological,kim2013surface}. This is expected because the saturation only disappears for Gd$^{3+}$ concentrations of $x \geq$ 0.03 \cite{kim2014topological,jiao2018magnetic}. Using the value of the resistivity at 4 K, we estimate the skin depth in a first approximation of the investigated samples to be $\delta _{D}$ = $\sqrt{\rho / \pi f \mu}$ $\approx$ 90 $\mu$m, 80 $\mu$m and 50 $\mu$m for $x$ = 0.0004, 0.004 and 0.02, respectively. These estimated values should be taken with extreme care as recent magnetotransport measurements have shown that subsurface cracks create additional conduction channels in SmB$_{6}$, which affects the maximum resistivity value obtained \cite{eo2019definitive}. A more dedicated investigation would be appropriate to compare resistivity values of our results with undoped samples. Consequently, probably we are underestimating the skin depth, which may be even larger than the sample size ($\approx$ 350 $\mu$m). Finally, it is important to notice that all samples show a similar gap in the resistivity, which agrees with previous Corbino disks measurements and the effects of Sm vacancies on the resistivity of SmB$_{6}$ \cite{eo2018robustness}.

The low-temperature specific heat $c_{p}$ is shown as a function of $T^{2}$ in the top panel inset of Fig. \ref{Fig1}. By extrapolating the high-$T$ $c_{p}$ data towards $T$ $\rightarrow$ 0 K, we estimate the Sommerfeld coefficients $\gamma$ $\sim$ 20 mJ/mol K$^{2}$, 30 mJ/mol K$^{2}$ and 35 mJ/mol K$^{2}$ for $x$ = 0.0004, 0.004 and 0.02, respectively. The red dashed line exemplifies the extrapolation for $x$ = 0.0004. These values should be taken with care, especially comparing samples with different Gd$^{3+}$ concentrations. It has been reported that undoped samples can show very distinct $\gamma$ values \cite{thomas2018quantum} and disorder could play a role in the Sommerfeld coefficient \cite{sen2018fragility}. More importantly, the low-temperature increase of $c_{p}$ is related to a Kondo-impurity-like behavior, which is similar to previous reports for Sm$_{1-x}$La$_{x}$B$_{6}$, Sm$_{1-x}$B$_{6}$ and Sm$_{1-x}$Gd$_{x}$B$_{6}$ \cite{fuhrman2018screened,orendac2017isosbestic}. Although this increase has been attributed to a local screening of the Gd$^{3+}$ ions \cite{fuhrman2018screened}, the local destruction of the Kondo lattice by a non-screened ion, which is a Kondo hole effect, also provides a reasonable explanation for this increase \cite{lawrence1996kondo}. We should note that it appears to be additional entropy into the system when we compare Sm$_{1-x}$La$_{x}$B$_{6}$ and Sm$_{1-x}$Gd$_{x}$B$_{6}$, which was reported before \cite{fuhrman2018screened,orendac2017isosbestic}. This additional entropy, for example, might arise from AFM exchange field fluctuations and/or a possible Gd$^{3+}$ interaction with local conduction electrons at ultra low-T ($\sim$ mK).

The magnetic susceptibility as a function of temperature is shown in the bottom panel of Fig. \ref{Fig1}. A clear $x$-dependent Curie-Weiss-like behavior is observed at low temperatures, whereas at higher temperatures the behavior of the magnetic susceptibility data is consistent with undoped SmB$_{6}$ \cite{lesseux2017anharmonic} (Fig. S1). Above $T$ $\approx$ 5 K the data starts to deviate from the Curie-like behavior, which means that the Gd$^{3+}$ contribution is not the most dominant one. We have used the low temperature region, $T \leq 4$ K, to isolate the contribution from the Gd$^{3+}$ ions itself, as shown in the inset of the bottom panel of Fig. \ref{Fig1}.

\begin{figure}[!ht]
\includegraphics[width=0.8\columnwidth]{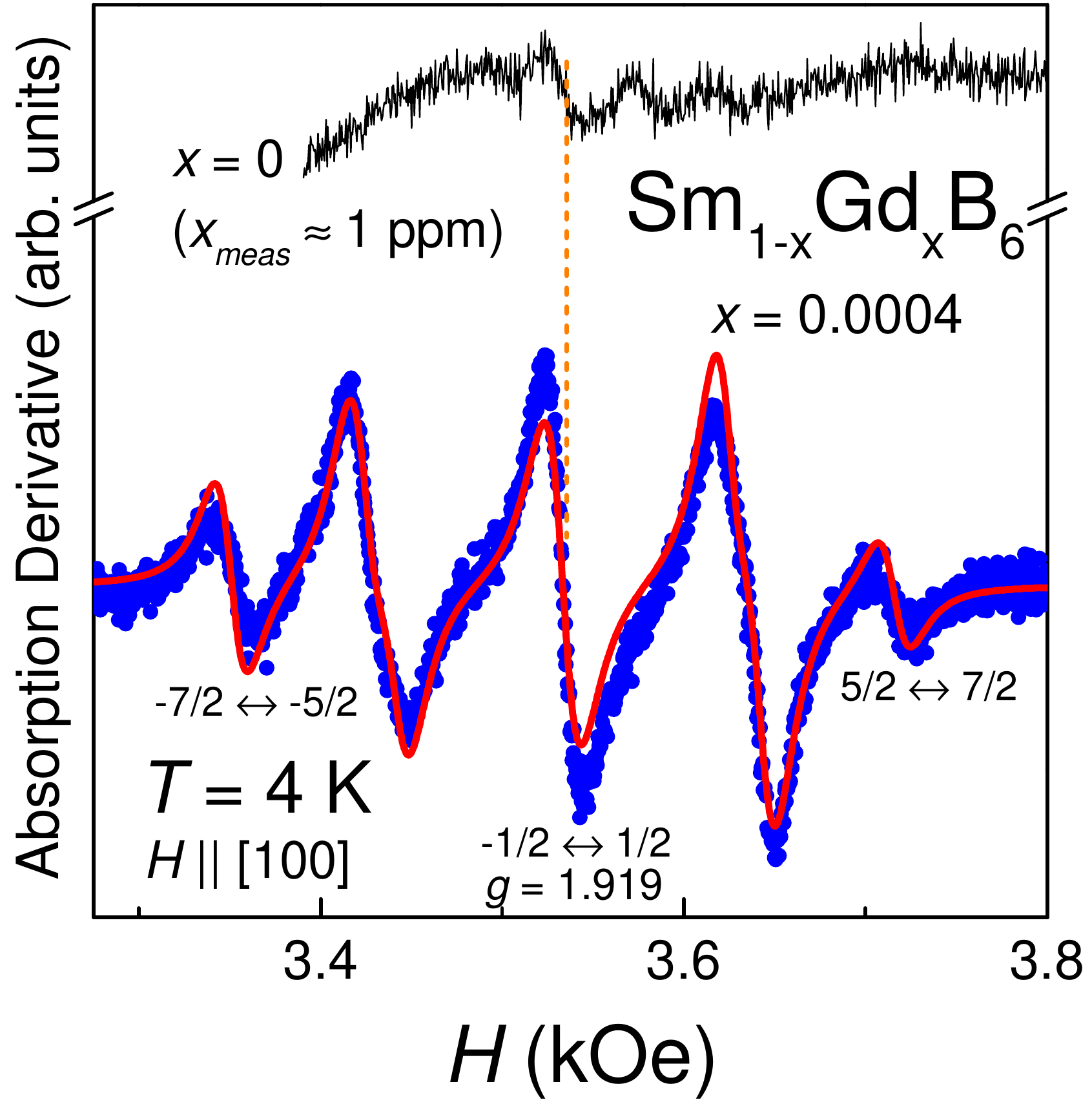}
\caption{X-band Gd$^{3+}$ ESR spectra for Sm$_{1-x}$Gd$_{x}$B$_{6}$ single crystals with $x$ = 0 and 0.0004. The field $H$ is applied parallel to the [100] direction. The dashed line shows the $g$-value of the -1/2 $\leftrightarrow$ 1/2 resonance. The red solid line is a simulation assuming a cubic insulator spin Hamiltonian with a CFP $b_{4}$ = - 9.5(3) Oe \cite{barnes1981theory}.}
\label{Fig2}
\end{figure}

In order to extract the magnetic moment of Gd$^{3+}$ and compared it to the expected values, we have used the measured concentration $x_{\mathrm{meas}}$ of our samples using ESR. In an ESR experiment we can estimate the number of resonant spins $N_{spins}$ comparing the ESR intensity $I$ with a well-known reference. The relationship between the sample $A$ ESR intensity and the $B$ reference can be written as \cite{abragam2012electron}

\begin{equation}
\frac{I_A}{I_B}=\frac{N_{spins}^A}{N_{spins}^B}  \left ( \frac{g_A}{g_B} \right )^2 \left [ \frac{S_A (S_A+1)}{S_B (S_B+1)} \right ] \left  ( \frac{T_B}{T_A} \right ) 
\label{Eq1}
\end{equation}
where $g$ is the $g$-value, $S$ the spin moment and $T$ the temperature. The ESR intensity $I$ is obtained double integrating our Gd$^{3+}$ ESR spectrum. The $g$-value is a parameter extracted from the Gd$^{3+}$ ESR resonance field, which is obtained fitting our Gd$^{3+}$ ESR spectrum using an admixture of absorption and dispersive derivatives. We will discuss the Gd$^{3+}$ ESR $g$-value in more details below. Fig. \ref{Fig2} shows the Gd$^{3+}$ ESR spectra for Sm$_{1-x}$Gd$_{x}$B$_{6}$ with $x$ = 0 and 0.0004 at 4 K. The Gd$^{3+}$ ESR spectrum for $x$ = 0.0004 is resolved, i.e., it contains seven distinct Lorentzian line shapes. This spectrum is the resonance of the Gd$^{3+}$ fine structure ($S$ = 7/2, selection rule $\Delta m$ = $\pm$ 1) and is characteristic of an insulating environment having a skin depth larger than the sample size. The red solid line is a simulation considering a cubic crystal field spin Hamiltonian with a crystal field parameter (CFP) $b_{4}$ = - 9.5(3) Oe \cite{barnes1981theory}. Even though the $\gamma$ values obtained are not the smallest reported in the literature for SmB$_{6}$ samples, it does not affect our results. The insulating Gd$^{3+}$ ESR line shape reinforces the clear insulating bulk environment in our samples. For the undoped compound we were able to observe a resonance with a very similar $g$-value compared with $x$ = 0.0004, as shown by the orange dotted line in Fig. \ref{Fig2}. This is an indication that this resonance could, most likely, be coming from Gd$^{3+}$ or Eu$^{2+}$ natural impurities in the undoped compound. Using Eq. \ref{Eq1}, the measured concentration $x_{\mathrm{meas}}$ estimated was $\approx$ 1 part per million, which is extremely low and shows the good quality of the undoped single crystals. We also did not observe an ESR signal from disordered Sm$^{3+}$ as reported by previous studies \cite{uemura1986electron,altshuler1999jahn}.

For the Gd$^{3+}$-doped compounds ESR spectra, as shown in Fig. \ref{Fig2}, using Eq. \ref{Eq1} we were able to estimate the actual concentrations of Gd$^{3+}$ based on $N_{spins}$ - $x_{\mathrm{meas}}$ = 0.00034(1), 0.0039(1) and 0.0189(1). Using these values, we fit the $\chi$ data for different concentrations with a Curie-Weiss law plus a Pauli $T$-independent term in the range of 2 K $\leq$ $T$ $\leq$ 4 K - see lower inset of Fig. \ref{Fig1}. The obtained magnetic moments are $\mu _{Gd}$ = 8.0(1) $\mu_{B}$/Gd, 8.2(1) $\mu_{B}$/Gd and 7.94(2) $\mu_{B}$/Gd for $x$ = 0.0004, 0.004 and 0.02, respectively. Thus, the full theoretically expected moments for Gd$^{3+}$ are observed. 

Our Gd$^{3+}$ ESR spectra as displayed in Fig. \ref{Fig2} do not support the scenario of a Gd$^{3+}$ dynamic screening. The ESR should be only observable for temperatures above the single impurity Kondo regime ($T$ $>>$ $T_{K}$), for which the Gd$^{3+}$ local moment is well defined \cite{baberschke1980esr}. Furthermore, we do not observe the ESR spin dynamics expected for a single impurity Kondo ion \cite{baberschke1980esr} or for a Kondo ion lattice \cite{schlottmann2018theory,kochelaev2017magnetic}. Therefore, these results do not corroborate the scenario of dynamic Gd$^{3+}$ screening proposed by Fuhrman $et$ $al.$ \cite{fuhrman2018screened}.

Besides proving information on spin dynamics, the Gd$^{3+}$ ESR line width $\Delta H$ also provides information on disorder and sample quality. For $x$ = 0.0004, the Gd$^{3+}$ resonance of the central line at $T$ = 4 K, which corresponds to the -1/2 $\leftrightarrow$ 1/2 transition, has a line width $\Delta H$ = 17(2) Oe. When the fine structure is collapsed into one resonance by varying the angle with respect to the applied magnetic field $H$ \cite{barnes1981theory,abragam2012electron}, we obtained $\Delta H$ = 19(2) Oe - fig. S2 a). This is consistent with samples known as good insulators \cite{buckmaster1972survey}. Further, the Gd$^{3+}$ line width at $T$ = 4 K has a similar value when compared with previous results for Er$^{3+}$-doped SmB$_{6}$ \cite{sturm1985esr,lesseux2017anharmonic}

\begin{figure}[tbp]
\includegraphics[width=0.8\columnwidth]{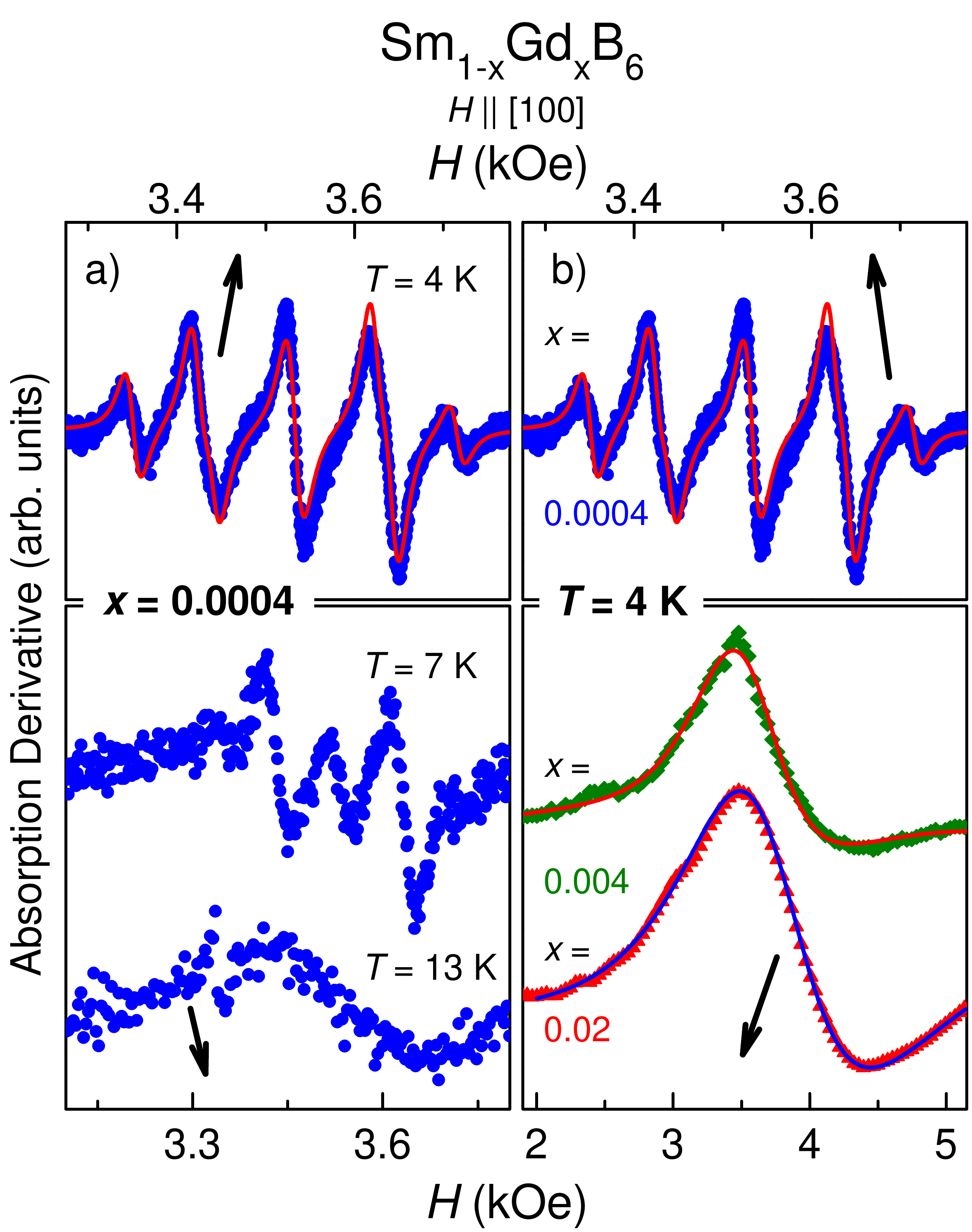}
\caption{Gd$^{3+}$ X-band ESR spectra a) $T$ and b) $x$ dependence for Sm$_{1-x}$Gd$_{x}$B$_{6}$, with $x$ indicated in the panels. The field $H$ is applied parallel to the [100] direction. The solid lines are the best fit obtained to the spectra using an admixture of absorption and dispersion derivatives for $x$ = 0.004 and 0.02 \cite{barnes1981theory}. For $x$ = 0.0004 the same fit as in Fig. \ref{Fig2} is shown.}
\label{Fig3}
\end{figure}

Figure \ref{Fig3} a) shows the Gd$^{3+}$ ESR line shape evolution as a function of temperature for Sm$_{1-x}$Gd$_{x}$B$_{6}$ for $x$ = 0.0004. As temperature increases, the seven lines merge into one line, which is a narrowing due to the interaction of the Gd$^{3+}$ 4$f$ local moments and the conduction electrons, known as exchange narrowing \cite{barnes1981theory,abragam2012electron,anderson1953exchange}. However, as it has been already shown for Kondo insulators \cite{garcia2011thermally,venegas2016collapse}, it is necessary to take into account the influence of the valence fluctuation of Sm$^{2.6+}$ at the Gd$^{3+}$ site to fully describe the narrowing effect.

The most intriguing change in the line shape occurs when, at constant temperature, the Gd$^{3+}$ concentration is increased from $x$ = 0.0004 to 0.004 and 0.02. As shown in in Fig. \ref{Fig3} b), the Gd$^{3+}$ ESR spectra at $T$ = 4 K for $x$ = 0.004 and 0.02 display a single asymmetrical Lorentzian line shape, also known as Dysonian line shape \cite{feher1955electron,dyson1955electron,barnes1981theory}, with no Gd$^{3+}$ fine structure. This line shape is characteristic of a metallic environment, where the skin depth is much smaller than the thickness of the sample. At the same temperature, the $x$ = 0.0004 sample still displays the seven distinct Lorentzians line shapes as expected for an insulator. Therefore, the local environment surrounding the Gd$^{3+}$ ions is changing as a function of the Gd$^{3+}$ concentration. At the same time this is not indicated in resistivity data, where the thermally activated bulk behavior still remains. As shown in the top panel of Fig. \ref{Fig1}, both the $T$-dependence of the resistivity data and skin depth are very similar for all concentrations, especially when we compare $x$ = 0.0004 and 0.004, showing a similar indirect gap and without any apparent change of the insulating ground state.

The remarkable change of the local Gd$^{3+}$ environment happens between $x$ = 0.0004 and 0.004. In order to understand the origin of such evolution, it is useful to compare it with nonmagnetic insulator Gd$^{3+}$ and Eu$^{2+}$ doped CaB$_{6}$ \cite{urbano2002different,wigger2004percolation,urbano2005gradual}, where the Kondo effect does not play a role.

In CaB$_{6}$, the local insulator-to-metal evolution as a function of Gd$^{3+}$ was interpreted as the introduction of an extra electron to the system, with Gd$^{3+}$-ions acting as electron donors and creating a hydrogen-like bound state within the gap. The bound states overlap and form a percolative network, i.e., there is a change of the ground state of the system. In that case, for higher Gd$^{3+}$ concentrations (0.003 $\leq$ $x$ $\leq$ 0.01), the ESR results were always consistent with the metallic resistivity data \cite{urbano2002different}. This is in contrast to the situation of Gd$^{3+}$-doped SmB$_{6}$, as stated above. The introduction of Eu$^{2+}$ in CaB$_{6}$ gives rise to a localized split-off bound state, with the Fermi energy lying in the gap of the semiconductor. This impurity state only percolates with $x$ $\geq$ 0.3, when the ground state becomes a ferromagnetic metal; however it is possible to observe an Eu$^{2+}$ Dysonian line shape for $x$ $\geq$ 0.15 \cite{wigger2004percolation,urbano2005gradual}.

SmB$_{6}$ is not expected to host a hydrogen-like doping mechanism \cite{rakoski2017understanding}. Furthermore, the non-parabolic band structure of SmB$_{6}$ leads to a significant increase of the calculated density of impurities required for percolation \cite{skinner2019properties}. Hence, the origin of the insulator-to-metal evolution in SmB$_{6}$ should differ from that in CaB$_{6}$. Our results agree with these two propositions. If the origin of the metallic islands and consequent evolution were due to a Gd$^{3+}$-induced hydrogen-like in-gap bound state, we should expect an evolution from insulating to a metallic resistivity, i.e., a change in the ground state of the compound, or a striking change in the hybridization gap value, which is not observed. 

Remarkably, the insulator-to-metal evolution is also noticed in the ESR response of Sm$_{1-y}$Eu$_{y}$B$_{6}$ samples. The Eu$^{2+}$ ESR spectrum of $y$ = 0.0004 shows a cubic insulator environment, with a well defined Eu$^{2+}$ hyperfine structure and a CFP $b_{4}$ $\approx$ - 50 Oe (fig. S2 b)). For higher concentrations, e.g. $y$ = 0.01, the Eu$^{2+}$ fine structure collapses and we obtain a Dysonian line shape. Previous reports have shown that a metallic antiferromagnetic ground state for Sm$_{1-y}$Eu$_{y}$B$_{6}$ only appears at or above $y = 0.4$, at least one order of magnitude higher than what it is observed by electron spin resonance \cite{yeo2012effects}. The Gd$^{3+}$ and Eu$^{2+}$-doped SmB$_{6}$ results suggest that there is not a percolation in the system, but a local effect which does not affect the global hybridization gap.

One alternative scenario is that the Gd$^{3+}$ ions are not screened by conduction electrons, although they occupy a Sm site. Instead of screening, these substitutions locally close the hybridization gap at the Gd$^{3+}$ site, increasing the local density of states and/or the local number of carriers. As already stated, this \lq\lq Kondo hole\rq\rq effect of Gd$^{3+}$ is consistent with the low-$T$ magnetic susceptibility and specific heat measurements shown in Fig. \ref{Fig1}, as well with the scaling proposed by Fuhrman $et$ $al.$ \cite{fuhrman2018screened}. Such effect could have two different local manifestations: the first one, and more unlikely due to the small amount of Gd$^{3+}$, resides in the possibility of local in-gap impurity states \cite{laurita2016anomalous,laurita2019impurities}. 

The natural scenario we propose is related to the formation of a bound state at the Gd$^{3+}$ site. This state should affect locally the hybridization gap around the Gd$^{3+}$ ions due to translational symmetry breaking. In other words, this substitution should create a spatial oscillation of the hybridization gap near the impurity \cite{figgins2011defects}. In the highly diluted case, such metallic islands are small and separated, in the sense that the Gd$^{3+}$ spins have a spin relaxation mechanism mediated by phonons. Hence we obtain a Gd$^{3+}$ ESR spectrum as expected for a cubic insulating environment. Based on previous theoretical reports \cite{figgins2011defects}, we propose that when there is quantum interference between such metallic islands, due to the increase of Gd$^{3+}$ concentration, effectively the islands grow in size and, as a consequence, the carriers will have mobility to enable metallic properties in the ESR line shape and ESR line parameters, which will lead to a local effect and will not reflect in the global properties of the system - such as $dc$ resistivity. Naturally a percolation should occur at higher concentrations. While for Gd$^{3+}$-doped SmB$_{6}$ we know that samples with $x$ = 0.05 still present an insulating behavior \cite{fuhrman2018screened}, La$^{3+}$-doped samples will show a metallic-insulating transition at x $\sim$ 0.3 \cite{orendac2017isosbestic}. Eu$^{2+}$-doped samples have a metallic-insulating transition at $x$ $\sim$ 0.4 \cite{yeo2012effects}.

Despite the local change of the environment, a more systematic in-depth theoretical investigation might need to introduce inhomogeneous effects, which may arise due to Gd$^{3+}$ ions in different magnetic states \cite{abragam2012electron}. Such effects may distort the ESR line shape asymmetry and line width. Although inhomogeneous effects should, in principle, play a role in fully describing the Gd$^{3+}$ ESR spectrum evolution, the possibility of new mechanisms should be a motivation for further theoretical works.

The intriguing question arising from the Kondo holes physics is related to the effects of metallic islands in a TKI. Would the metallic islands have a nontrivial topology? From an ESR viewpoint, the fingerprint of a nontrivial topological state could be related to the appearance of an ESR diffusive-like line shape for the signal of the ESR local moment probe \cite{lesseux2016unusual}. However, one of the main ingredients for such unusual effect is a phonon-bottleneck regime, which is not present in these studied Gd$^{3+}$ concentrations. Therefore, we cannot extract any information about the topology of the system from an ESR perspective for the studied samples. Further experiments exploring different Gd$^{3+}$ concentrations could be valuable in order to clarify this pressing question from our results.

\begin{figure}[tbp]
\includegraphics[width=0.8\columnwidth]{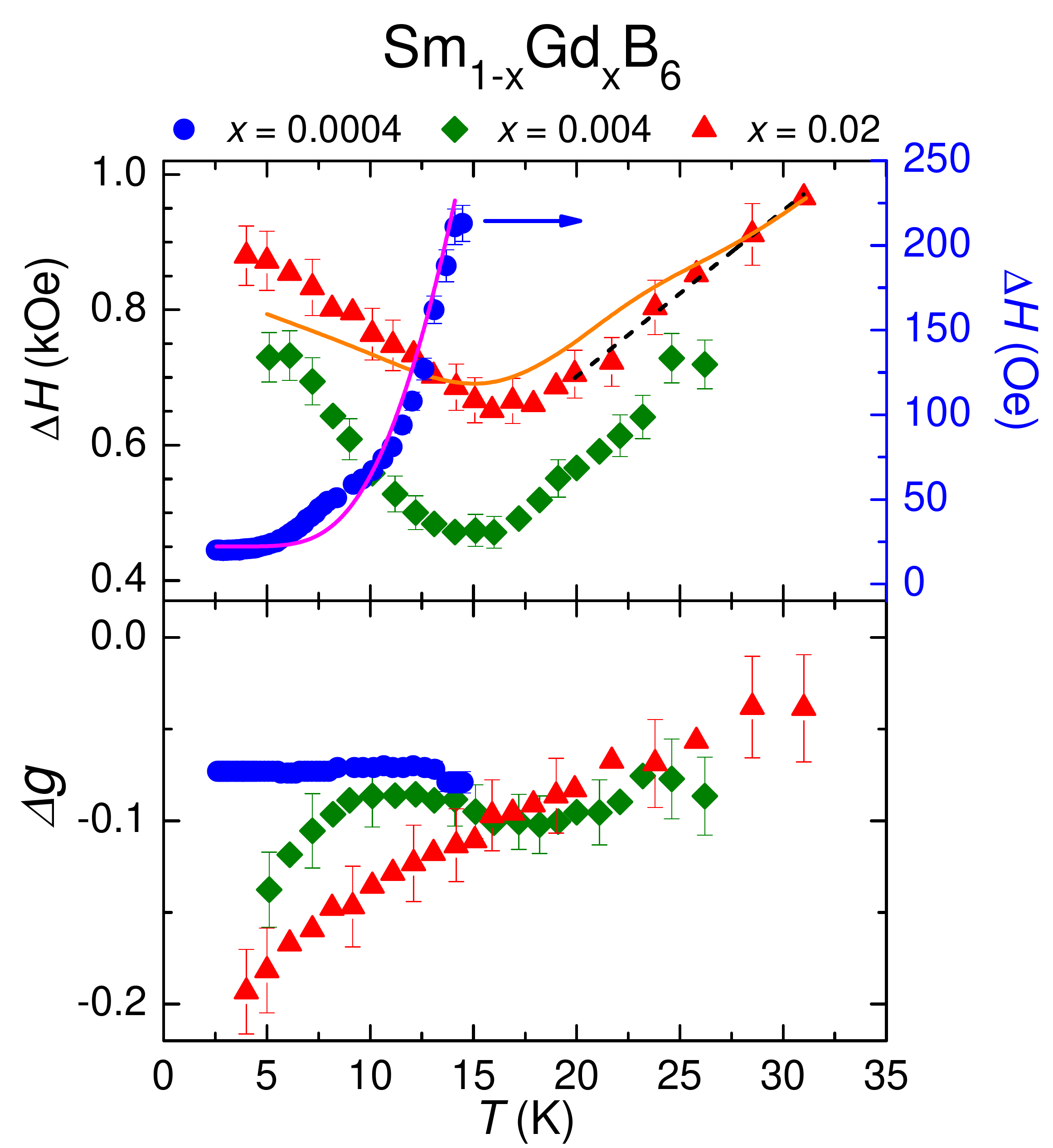}
\caption{Top panel: $T$-dependence of the Gd$^{3+}$ ESR line width $\Delta$$H$ for Sm$_{1-x}$Gd$_{x}$B$_{6}$ compound with $x$ = 0.0004, 0.004 and 0.02. While the external magnetic field was applied into the [100] direction for $x$ = 0.004 and 0.02, for $x$ = 0.0004 the magnetic field was applied 30 degrees from the [100] direction in the (110) plane (collapsed ESR spectrum). The solid lines are fits explained in the text. The dotted line is a guide to the eye. Bottom panel: $\Delta g$ as a function of temperature.}
\label{Fig4}
\end{figure}

In ESR, the line width $\Delta H$ is proportional to $1/T_{2}$, where $T_{2}$ is the spin-spin relaxation time. Hence, using an admixture of absorption and dispersive derivatives to fit the Gd$^{3+}$ ESR line shapes, we can extract $\Delta H$ and obtain information about the interaction between the Gd$^{3+}$ 4$f$ local moments and their environment, i.e., we can extract the ESR spin dynamics. The solid lines in Fig. \ref{Fig3} b) indicate the fit for $x$ = 0.004 and 0.02. In order to exclude crystal field effects, the $T$-dependence of $\Delta H$ for $x$ = 0.0004 was obtained when the Gd$^{3+}$ ESR spectrum is collapsed into one resonance \cite{abragam2012electron}. The top panel of Fig. \ref{Fig4} shows the evolution of the line width $\Delta H$ as a function of temperature. 

For the highly diluted $x$ = 0.0004 sample, it is possible to observe an exponential dependence of the Gd$^{3+}$ ESR line width as a function of temperature. The interconfigurational fluctuation model (ICF) \cite{hirst1970theory,gambke1978epr,venegas1992epr} shows that the fluctuation between the 4$f^{n}$ and 4$f^{n+1}$ configurations of the Sm ions provokes a fluctuating field at the Gd$^{3+}$ site, which explains the exponential increase of $\Delta H$

\begin{equation}
\Delta H = \Delta H_{0} + bT + Ae^{-E_{ex}/T},
\end{equation} 
where $\Delta H_{0}$ is the residual Gd$^{3+}$ line width, $b$ the Korringa relaxation rate \cite{barnes1981theory,abragam2012electron}, $E_{ex}$ the excitation energy of the Sm ions and $A$ a constant. The Korringa relaxation $b$ is related to the spin-flip scattering relaxation mechanism which arises from the interaction between conduction electrons and the Gd$^{3+}$ 4$f$ local moments \cite{barnes1981theory,abragam2012electron}. We obtain $b$ = 0 Oe/K and $E_{ex}$ = 56 K (magenta solid line), which is, notably, of the order of the hybridization gap \cite{eo2018robustness}. This result is another microscopic hint that Gd$^{3+}$ ions are not introducing extrinsic charge in-gap states when substituting Sm ions in SmB$_{6}$.

Comparing the line width $\Delta H$ for all samples, we can see a clear difference between the highly dilute system ($x$ = 0.0004) and the other two Gd$^{3+}$ concentrations. The relaxation changes dramatically when the Gd$^{3+}$ sites start to interact, which can be verified in trying to use the same ICF model for $x$ = 0.02 (orange solid line). In this case, considering a phenomenological exchange interaction between Gd$^{3+}$ sites with an internal exchange field $H_{ex}$ = 20 Oe \cite{gambke1978epr}, we obtain a Korringa relaxation rate $b$ = 1.5(5) Oe/K and $E_{ex}$ = 110(10) K which is twice larger than those for $x$ = 0.0004. This clearly indicates that the model does not necessarily apply for higher Gd$^{3+}$-concentrations, where the correlation between the metallic islands covers the effect of interconfigurational fluctuations.

As already mentioned, from the admixture of absorption and dispersive derivatives fitting we can also extract the Gd$^{3+}$ resonance field $H_{0}$. Consequently we obtain the Gd$^{3+}$ experimental $g$-value $g_{exp}$ = $h\nu/\mu_{B} H_{0}$, where $h$ is the Planck constant and $\mu_{B}$ the Bohr magneton. The bottom panel of Figure \ref{Fig4} shows the temperature dependence of Gd$^{3+}$ $\Delta g$ = $g_{exp}$ - $g_{insulator}$, where $g_{insulator}$ = 1.993(1) for an isolated Gd$^{3+}$ ion \cite{abragam2012electron}. In metals, for the simplest scenario \cite{rettori1997esr,duque2009magnetic,pagliuso1999electron}, the $g$-shift $\Delta g$ is given by $\Delta g$ = $J _{fs}$$\eta(E_{F})$ \cite{barnes1981theory}, where $J_{fs}$ is the effective exchange interaction between the $4f$ local moments and the conduction electrons and $\eta (E_{F})$ the DOS at the Fermi level for one spin direction. Positive $g$-shifts are expected due to the ferromagnetic (atomic) interaction between the 4$f$ local moments and $s$ and/or $d$ conduction electrons. Negative $g$-shifts are obtained when the contribution at the Fermi level is coming from $p$ and/or $f$ conduction electrons, since their magnetic interaction with the 4$f$ local moments occurs via a virtual bound state \cite{tao1971hyperfine} and, consequently, an antiferromagnetic (covalent) exchange interaction appears \cite{barnes1981theory,tao1971hyperfine}. In the simplest scenario for metals it is crucial to note that the Korringa relaxation $b$, which is extracted from $\Delta H$ $T$-dependence, is also proportional to the exchange interaction between the 4$f$ local moments and the conduction electrons. In other words, $b$ $\propto$ $J ^{2} _{fs}$$\eta^{2}(E_{F})$ when we neglect $\textbf{q}$-dependence or multiband effects \cite{barnes1981theory}. Another crucial point is that internal ferromagnetic or antiferromagnetic fields can also shift the resonance field, generating apparent positive or negative $g$-shifts.

The $\Delta g$ = - 0.074(2) for $x$ = 0.0004 is $T$-independent. Since the Korringa relaxation at this concentration is negligible and the spin relaxation is phonon-mediated, this negative value most likely is due to the interaction between the Gd$^{3+}$ and Sm$^{2.6+}$ 4$f$ electrons. In other words, our results indicate that there is no relevant overlap between the Gd$^{3+}$ 4$f$ and the conduction electrons wave functions at these investigated temperatures, which is in agreement with our scenario. Since the magnetic contribution of the Sm$^{2.6+}$ is $T$-independent at this temperature range, it behaves like a constant AFM internal field at the Gd$^{3+}$ site, which is consistent with recent magnetic dichroism results \cite{fuhrman2019magnetic}. In this scenario, the AFM net local field at the Gd$^{3+}$ site would be $H^{net}$ = 130(5) Oe. 

For the higher concentration samples there is a clear decrease of the Gd$^{3+}$ $\Delta g$ towards low temperatures. This is an indication of an exchange interaction between the Gd$^{3+}$ sites, which is expected to be negative since GdB$_{6}$ is an antiferromagnetic compound with T$_{N}$ $\approx$ 15 K \cite{kunii1985electronic}. The possibility of Gd-Gd interactions for such small Gd$^{3+}$-concentrations ($x$ = 0.004) is also suggested by STS studies which show that defects in SmB$_{6}$ have a \lq\lq healing length\rq\rq  of $\sim$ 2 nm \cite{jiao2018magnetic}. If we assume that this would also characterize the bulk and not just the surface, the affected volume around a defect could be at the order of $\sim$ 500 unit cells. It is worth noting that Gd-Gd interaction for $x$ = 0.004 and 0.02 is also reflected in the low-$T$ increase of the linewidth, as shown in the top panel of Fig. \ref{Fig4}, although this increase does not scale with the Gd$^{3+}$ concentration.

As discussed above, due to the correlation between the metallic regions, the local density of carriers may become mobile and start to play a role in the relaxation mechanism, resulting in a finite value of the Korringa relaxation rate even at lower temperatures for $x$ = 0.004 and 0.02. As a consequence, the Gd$^{3+}$ ESR turns into a single Dysonian lineshape with a small Korringa relaxation, i.e., the environment changes from an insulator-like to a metallic-like. Although these local effects can be seen, the global hybridization gap is not affected, as shown by $dc$ resistivity data.

This ESR scenario helps to understand the apparent discrepancy between $ac$ and $dc$ conductivity data \cite{eo2018robustness,gorshunov1999low,laurita2016anomalous,laurita2019impurities,eo2019definitive}. The local response of the metallic islands will not affect the global hybridization gap due to the lack of percolation, hence at the $dc$ limit it is only possible to observe an insulating behavior. Recent theoretical models argue that such localized states could produce quantum oscillations in the magnetization, which is a plausible way to understand the presence of quantum oscillations in magnetization and absence in resistivity \cite{tan2015unconventional,eo2019definitive}. Defects, such as Sm vacancies and natural magnetic impurities, are responsible for localized metallic islands, which allegedly could produce such oscillations \cite{fu2018quantum}. Although recent theoretical results indicate that the reported oscillations could be supported by the bulk band structure of this compound \cite{zhang2020understanding}, our results are also consistent with the emergence of quantum oscillations from bulk impurities \cite{fu2018quantum}. In contrast to this theoretical proposition \cite{zhang2020understanding}, we should note that SmB$_{6}$ does not show a magnetic ordered state at ambient pressure \cite{barla2005high,zhou2020hall}. Finally, the Kondo hole scenario for Gd$^{3+}$ is also valid for Eu$^{2+}$-doped SmB$_{6}$ and is a plausible way to explain the scaling as a function of Gd$^{3+}$ concentration in SmB$_{6}$ described by Fuhrman $et$ $al.$ \cite{lawrence1996kondo,fuhrman2018screened}.

Our results show that there is a local insulator-to-metal evolution in Gd$^{3+}$-doped SmB$_{6}$ at very low concentrations. This observation was possible only due to the fact that we are directly probing the Gd$^{3+}$ in the bulk locally. 

\section{\label{sec:conclusion}IV. Conclusion}

In summary, we performed electron spin resonance and complementary macroscopic measurements in the Kondo insulator Sm$_{1-x}$Gd$_{x}$B$_{6}$ with $x$ = 0, 0.0004, 0.004 and 0.02. The Gd$^{3+}$ ESR spectra at 4 K for different concentrations showed two clearly different behaviors. For $x$ = 0.0004 we observed an insulator-like Gd$^{3+}$ ESR line shape, while for $x$ = 0.004 and 0.02 we obtained a \lq\lq Dysonian\rq\rq metallic line shape, characteristic of a conductive environment. Hence, the hybridization gap at the Gd$^{3+}$ site collapses, which can result in an effective formation of larger metallic islands as a function of  magnetic impurity concentration. This scenario is consistent with the observed ESR spin dynamics and shift of ESR $g$-factors. These localized states explain the discrepancy between $dc$ and $ac$ conductivity measurements, and can also provide a different point-of-view regarding quantum oscillations in this system. Further experiments, such as ESR in \lq\lq neutral\rq\rq doping samples, e.g. Gd$^{3+}$-Sr$^{2+}$ and Eu$^{2+}$-La$^{3+}$ dopings, and NMR measurements in Gd$^{3+}$-doped SmB$_{6}$ could be valuable to understand the evolution of the metallic islands in more details.

\begin{acknowledgments}

We acknowledge constructive discussions with C. Rettori and S. M. Thomas. This work was supported by FAPESP\ (SP-Brazil) Grants No 2018/11364-7, 2017/10581-1, 2016/14436-3, 2013/17427-7, 2012/04870-7, 2012/05903-6, CNPq Grants No 309483/2018-2, 442230/2014-1 and 304649/2013-9, CAPES and FINEP-Brazil. ZF acknowledges the funding from NSF-1708199. Work at Los Alamos National Laboratory (LANL) was performed under the auspices of the U.S. Department of Energy, Office of Basic Energy Sciences, Division of Materials Science and Engineering.

\end{acknowledgments}

\end{document}


\title{Supplementary Information:\\
Metallic islands in the Kondo insulator SmB$_{6}$}

\author{J. C. Souza$^{1,2}$, P. F. S. Rosa$^{3}$, J. Sichelschmidt$^{2}$, M. Carlone$^{4}$, P. A. Venegas$^{5}$, M. O. Malcolms$^{1}$, P. M. Menegasso$^{1}$, R. R. Urbano$^{1}$, Z. Fisk$^{6}$, and P. G. Pagliuso$^{1}$}

\affiliation{$^{1}$Instituto de F\'isica \lq\lq Gleb Wataghin\rq\rq,
UNICAMP, 13083-859, Campinas, SP, Brazil\\
$^{2}$Max Planck Institute for Chemical Physics of Solids, D-01187 Dresden, Germany\\
$^{3}$Los Alamos National Laboratory, Los Alamos, New Mexico 87545, USA\\
$^{4}$Departamento de F\'isica, Faculdade de Ci\^encias, UNESP, C.P. 473, 17033-360, Bauru, SP, Brazil\\
$^{5}$POSMAT, Faculdade de Ci\^encias, UNESP, C.P. 473, 17033-360, Bauru, SP, Brazil\\
$^{6}$Department of Physics and Astronomy, University of California, Irvine, California 92697,USA}



\maketitle
\section{\label{sec:magnetization}I. Magnetization measurements}

\begin{figure}[!ht]
\includegraphics[width=0.95\columnwidth]{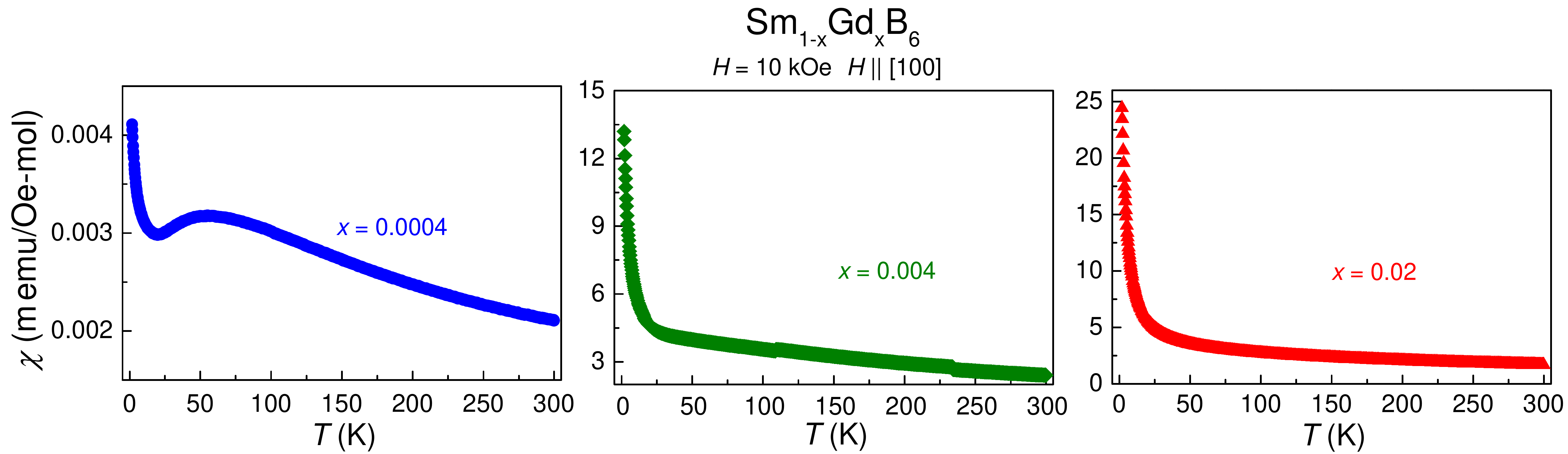}
\caption{Magnetic susceptibility as a function of temperature for Sm$_{1-x}$Gd$_{x}$B$_{6}$ for $x$ = 0.0004, 0.004 and 0.02, respectively.}
\label{FigS1}
\end{figure}

Figure S1 shows in more details the magnetic susceptibility as a function of temperature for Sm$_{1-x}$Gd$_{x}$B$_{6}$ for $x$ = 0.0004, 0.004 and 0.02, respectively. At low temperatures it is possible to observe a Curie-Weiss tail, suggesting that the Gd$^{3+}$ ions are free paramagnetic impurities embedded in the Kondo lattice. From a Curie-Weiss fit we also obtained the Curie-Weiss temperatures $\theta_{CW}$ = - 1(5) K, - 2(5) K and - 3(5) K for $x$ = 0.0004, 0.004 and 0.02, respectively. Similar results were reported for Nd$^{3+}$-doped CeRhIn$_{5}$ \cite{rosa2016unusual}.

\newpage
\maketitle
\section{\label{sec:ESR}II. ESR experiments}

\begin{figure}[!ht]
\includegraphics[width=0.95\columnwidth]{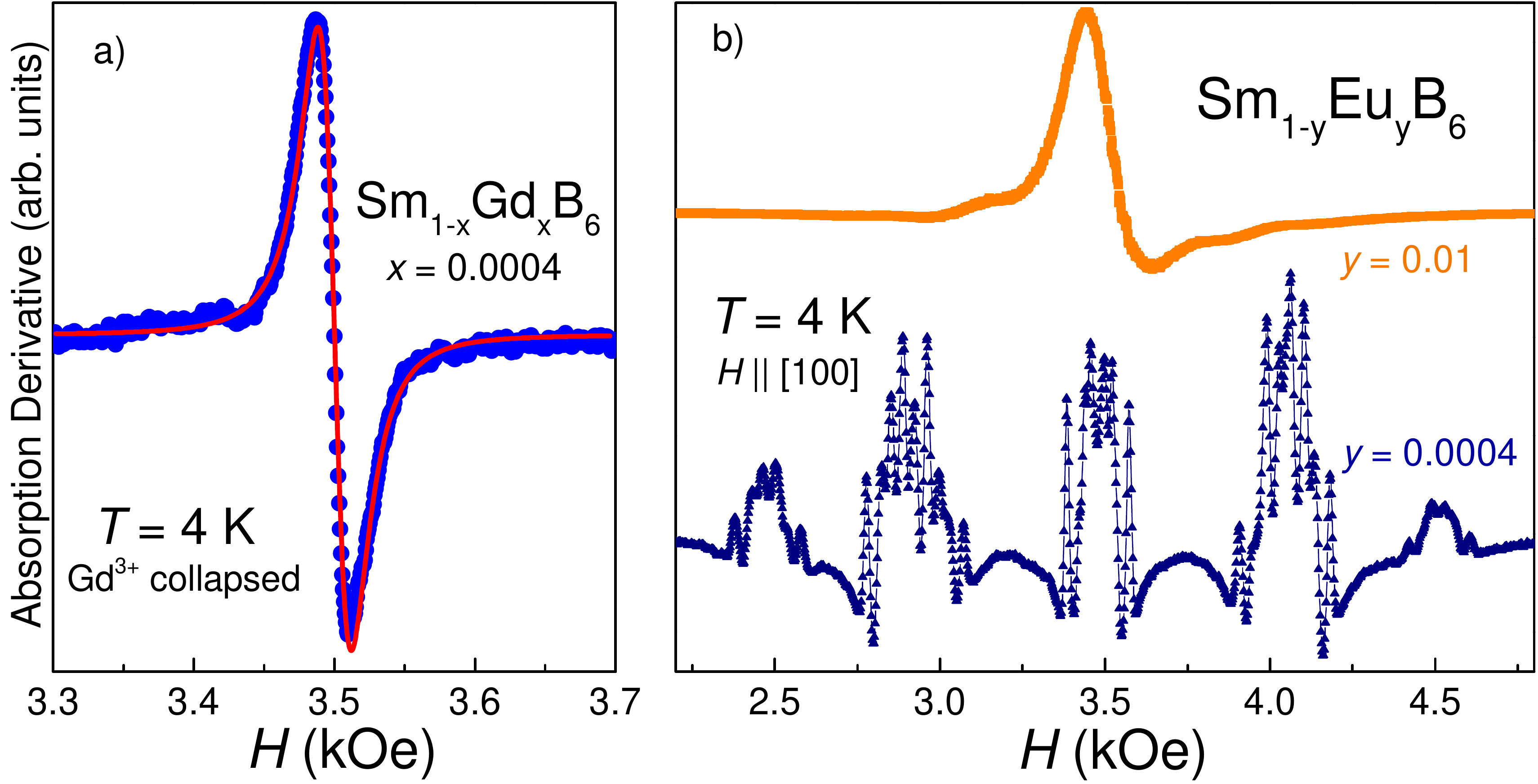}
\caption{a) Gd$^{3+}$ ESR spectrum at $T$ = 4 K for the magnetic field applied at 30 degrees from the [001] direction in the (110) plane. b) Eu$^{2+}$ ESR spectra at $T$ = 4 K for the magnetic field applied parallel to the [100] direction.}
\label{FigS2}
\end{figure}

Figure S2 a) shows the collapsed Gd$^{3+}$ ESR spectrum at $T$ = 4 K for Sm$_{0.9996}$Gd$_{0.0004}$B$_{6}$. The solid line is the best fit obtained using an admixture of absorption and dispersion derivatives. The obtained parameters were $\Delta H$ = 19(2) Oe and $g$ = 1.920(2).

Figure S2 b) shows the Eu$^{2+}$ ESR spectra evolution for Sm$_{1-y}$Eu$_{y}$B$_{6}$ for $y$ = 0.0004 and 0.01. For $y$ = 0.0004 it is possible to observe the Eu$^{2+}$ fine structure and the additional lines are related to hyperfine resonance due to the $^{151}$Eu$^{2+}$ ($I$ = 5/2) and $^{153}$Eu$^{2+}$ ($I$ = 5/2) isotopes. For $y$ = 0.01 it is possible to observe a single asymmetrical Eu$^{2+}$ Lorentzian line shape, typical for metallic environments.